\documentstyle [12pt] {article}

\newcommand{\be}{\begin{equation}}
\newcommand{\ee}{\end{equation}}
\newcommand{\ba}{\begin{array}}
\newcommand{\ea}{\end{array}}
\newcommand{\bea}{\begin{eqnarray}}
\newcommand{\eea}{\end{eqnarray}}
\newcommand{\tr}{{\rm tr}}
\newcommand{\Tr}{{\rm Tr}}
\newcommand{\vier}{\\ [4 pt]}
\newcommand{\acht}{\\ [8 pt]}

\begin{document}
\begin{titlepage}
\title{
\hfill {\normalsize UPR--0567T} \\
\hfill \\ \hfill \\
Anomaly Cancellation in Six
Dimensions\thanks{Work supported by Deutsche Forschungsgemeinschaft.}}

\vspace{5.cm}
\author{
Jens Erler \\ \hfill \\
{\small David Rittenhouse Laboratory} \\
{\small University of Pennsylvania} \\
{\small Philadelphia, PA 19104 (USA)}}
\date{\quad}
\maketitle
\begin{abstract}
\noindent
I show that anomaly cancellation conditions are sufficient to determine the
two most important topological numbers relevant for
Calabi-Yau compactification
to six dimensions. This reflects the fact that K3 is the only non-trivial
CY manifold in two complex dimensions. I explicitly construct the
Green-Schwarz counterterms and derive sum rules for charges of additional
enhanced U(1) factors and compare the results with all possible Abelian
orbifold constructions of K3. This includes asymmetric orbifolds as well,
showing that it is possible to regain a geometrical interpretation for this
class of models. Finally, I discuss some models with a broken $E_7$ gauge group
which will be useful for more phenomenological applications.
\end{abstract}
\thispagestyle{empty}
\end{titlepage}

\setcounter{page}{1}

\section{Introduction}
It has been clear that anomaly cancellation belongs to the outstanding
properties of superstrings, ever since Green and Schwarz~\cite{GS} showed
that the zero slope limit of the SO(32) superstring theory gives rise to an
anomaly free $D=10$ supergravity theory and evades the no-go theorem of
Alvarez-Gaum\'e and Witten~\cite{AW}.
On general grounds~\cite{tHooft,AW} this property persists
in any approximation of the theory and in particular in an effective
supergravity theory in a lower number of dimensions. Such a
theory can be obtained by {\em direct\/} construction of strings in $D<10$
or by {\em geometrical\/} compac\-ti\-fi\-cations on Calabi-Yau
manifolds~\cite{CHSW}.

In this paper I reverse the direction and ask to what extent
the conditions of absence of anomalies can give us information about
the existence of Calabi-Yau (CY) manifolds. This is in the spirit of
a paper by Seiberg~\cite{Seiberg},
where the number of moduli fields was derived using the
absence of anomalies in type IIB supergravity in six dimensions.
Here I discuss the
heterotic string and show in section~\ref{ac} that the requirements lea\-ding
to a six-dimensional supergravity theory which is free of gauge and
gravitational anomalies turn out to be sufficient to determine the
main topological data of Kummer's third surface (K3). These are the number
of independent (1,1)-forms ($h_{1,1}$) and of (0,1) forms with values in the
endomorphisms of the tangent bundle, i.e.\ ${\rm dim} H^1 (End\, T)$.
In physical terms, they determine the number of generations
transforming in the {\bf 56} of $E_7$ and the number of singlet fields.
In section~\ref{ac} I also derive the explicit Green-Schwarz
counterterms for six-dimensional superstring models as well as the
transformation rules for the antisymmetric tensor field. Again, this is done
by using arguments related to anomalies only.

Of course, it would be of great importance to obtain a similar
result for the much more complicated case of four-dimensional theories.
Here a huge number of CY manifolds could be constructed~\cite{KS},
but the complete classification is still an open problem. At the same time
the absence of gauge anomalies in $D=4$ supergravity theories
can be achieved for an arbitrary number of generations
transforming in the ${\bf 27}$ or $\overline{\bf 27}$ of $E_6$.
Thus, a straightforward
generalization of the $D=6$ case is not possible. On the other hand,
considerations of other kinds of anomalies might help to improve the situation.
However, this is outside the scope of this paper and the four-dimensional case
must be treated elsewhere.

In section~\ref{orbifold} I find all possible $Z_N$
orbifold limits of K3. In particular, by noticing the uniqueness of K3, it
becomes clear that even for asymmetric orbifolds~\cite{NSV} we regain an
unambiguous geometrical interpretation\footnote{Here I assume Gepner's
conjecture~\cite{Gepner} to be true, which states that all (4,4) supersymmetric
conformal field theories correspond to $\sigma$-models on K3.}.
In contrast to symmetric or\-bi\-folds~\cite{DHVW,IMNQ,EK}, these models
{\em cannot\/} be obtained by using
identical shift vectors acting in the gauge lattice and in the lattice of
bosonized NSR fermions. In this publication I introduce this type of
construction and point out the similarities to and differences from the
symmetric case. Compactifications to four dimensions will be treated
in~\cite{asym}, where I will present the resulting (2,2) models.

The symmetric orbifold limits of K3 were already discussed in~\cite{Walton}.
However, the computation of singlet fields in~\cite{Walton} is incorrect
for non-prime $Z_N$ orbifolds. In fact, this determination is somewhat involved
and special care is needed for the projection onto twist invariant states.
Fortunately, the anomaly cancellation conditions of section~\ref{ac}
supply us with a perfect check. Moreover, they give us a number of sum rules
for charges under the additional $U(1)$'s, which serve as additional tests.

Restrictions and relations coming from anomaly cancellation, which are
often related to index theorems, are also useful for more phenomenological
questions. For the heterotic string to be of any relevance to describe
nature, almost all gauge symmetries of the 496-dimensional gauge group in
$D=10$ must be broken at low energies. Significant progress has been made in
the construction of vacuum configurations coming very close to what
we find phenomenologically. Here it is of particular importance to break
the rank of the gauge group. This is usually achieved by a variety of
mechanisms, which all have in common the phenomenon that vacuum expectation
values (vev's) are given to scalar (moduli) fields connecting degenerate
string vacua.
Presently we do not understand the dynamics which determines the values of the
moduli, but it is evident that it has to be very efficient. Therefore it
appears quite remarkable that an $SU(3) \times U(1)$ subgroup remains unbroken.
If the dynamically preferred values of moduli do not correspond to
enhanced gauge symmetries, which is also suggested by discussions
of non-pertubative potentials~\cite{FT}, why then, is the gauge symmetry not
completely broken? A possible explanation would be that there is no
moduli direction to break $SU(3) \times U(1)$. The reason for this in turn,
should then be some kind of index. Merely imposing some global symmetry would
not improve the situation, since in general the moduli vev's would break it
as well.

I illustrate the above mentioned efficiency of
symmetry breaking by means of a simple example using continuous Wilson lines
in the $Z_3$ orbifold.

In appendix~\ref{gt}, I present group theoretical identities
relevant for anomaly cancellation in four and six dimensions.

\section{Anomaly cancellation in six dimensions}
\label{ac}

In this section, I discuss some facts about anomaly cancellation in
six-dimensional supergravity theories. The basic diagram to be examined
is the box with an even number of external gravitons and gauge
fields~\cite{AW}.
The resulting anomalous diagrams can be classified as purely
gravitational, purely gauge or mixed gauge and gravitational. The pure
gauge anomalies will be referred to as {\em quartic\/} if all external
gauge fields belong to one group factor, and as {\em cubic\/} if three
gauge fields belong to one group factor and the fourth one to a $U(1)$.

For the theory to be anomaly free, one of two conditions have to be met
for each type of anomaly.
Either the coefficient of the respective anomaly vanishes or the one loop
anomaly can be cancelled by the variation of a Green-Schwarz tree-level
counterterm~\cite{GS} involving an antisymmetric tensor field. The latter
option requires a peculiar factorization of the non-vanishing anomaly
into two expressions each quadratic in the gravitational ($R$) resp.\
gauge field strengths ($F$).

Suppose now some non-Abelian factor $G_A$ of the gauge group possesses
a fourth-order Casimir invariant. Then the quartic gauge anomaly
is of the form
\be \label{quartic}
   \mu_A \tr F_A^4 + \nu_A (\tr F_A^2)^2 .
\ee
Obviously, the above mentioned factorization is impossible
unless all coefficients $\mu_A$ vanish. Likewise, the part of the pure
gravitational anomaly which cannot be written as a square necessarily has to
vanish.

If the gauge group contains a non-Abelian factor for which a third-order
invariant exists {\em and\/} at least one Abelian factor, then a cubic anomaly
is possible. Again it is necessary that it vanishes. As is well known from
anomaly considerations in four dimensions, only $SU(N)$ groups
with $N \geq 3$ have third order invariants. However, there is another case
where we can meet a cubic anomaly. This happens if we have at least two
$U(1)$ factors and consider the diagram with three photons belonging
to one $U(1)$ and the remaining photon to the other one. In general this
kind of cubic anomaly does not vanish. Once complete factorization is
achieved, however, it is always possible to find linear combinations of the
$U(1)$'s such that all cubic anomalies vanish. This then also ensures that
anomalies containing gauge fields of three or four different group factors
vanish.

The total anomaly originates from four sources~\cite{AW,GSWest,GSW},
\be
   I=I_{3/2}(R) - I_{1/2}(R) + I_{1/2}(R,F) - \sum_{i} s^i I_{1/2}^i(R,F) ,
\ee
where the first is due to the gravitino, the second to the dilatino, the
third to gauginos and the last to matter
fermions\footnote{There is also a self-dual antisymmetric tensor field
in the supergravity multiplet. Its contribution to the anomaly, however,
is cancelled by the anti-self-dual tensor field in the dilaton
multiplet. See also the discussion following equation~(\ref{SON}).}.
The dilatino and matter
fermions contribute with a minus sign, since supersymmetry requires them
to have opposite chirality as compared to the gravitino and gauginos.
The factor $s^i$ counts the multiplicity of representation $R_i$.
Using results from~\cite{AW,GSW} we find
\be \label{anomaly}
\ba{lll}
        i (2\pi)^3 I &=& {1\over 5760}(244+y-s) \tr R^4 \vier
   &+& {1\over 4608}(-44+y-s) (\tr R^2)^2 \vier
   &-& {1\over 96} \tr R^2
      [\Tr F^2_A  - \sum\limits_{i,A} s^i_A (\tr_{R^i} F^2_A)] \vier
   &+& {1\over 24} [\Tr F^4_A
       - \sum\limits_{i,A} s^i_A (\tr_{R^i} F^4_A)] \vier
   &-& {1\over 6} \sum\limits_{i,j,A,B} s_{AB}^{ij}
      (\tr_{R^i} F^3_A) (q_B^j F_B) \vier
   &-& {1\over 4} \sum\limits_{i,j,A,B} s_{AB}^{ij}
     (\tr_{R^i} F^2_A)(\tr_{R^j} F^2_B) \vier
   &-& {1\over 2} \sum\limits_{i,j,k,A,B,C} s_{ABC}^{ijk}
     (\tr_{R^i} F^2_A) (q_B^j F_B) (q_C^k F_C) \vier
   &-& \sum\limits_{i,j,k,l,A,B,C,D} s_{ABCD}^{ijkl}
     (q_A^i F_A) (q_B^j F_B) (q_C^k F_C) (q_D^l F_D).
\ea
\ee
$y$ denotes the dimension of the total gauge group, $s$ the total number
of hypermultiplets (matter), $s^i_A$ the number of hypermultiplets
transforming in representation $R^i$ of group factor $G_A$, $s_{AB}^{ij}$
the number of hypermultiplets transforming in representation
$(R^i,R^j)$ under $G_A \times G_B$, etc. $\Tr$ refers to the trace in
the respective adjoint representation and $q_A$ to the charge under $U(1)_A$.
The trace over curvature matrices in $R$ is in the vector representation of
$SO(5,1)$.
Notice the different prefactors in front of the various pure gauge terms.
They arise because a different number of gauge group
factors are involved yielding different combinatorical factors. The last two
terms can only arise if there are at least two or four $U(1)$ factors,
respectively.

{}From the previous discussion, it is now clear that the coefficient of
the first term must vanish. Thus, as already found
in~\cite{RSSS,GSWest,Walton}, we have to require
\be
\label{OHR}
   s - y = 244.
\ee
It is important to emphasize that relation~(\ref{OHR}) has to be strictly
satisfied for any \mbox{$N=1$} superstring vacuum in six dimensions and in
particular for any orbifold model\footnote{In~\cite{Walton} some of
the presented orbifold spectra fail to satisfy~(\ref{OHR}) and consequently
{\em cannot\/} be correct.}. An immediate application of~(\ref{OHR}) is its
interpretation as a {\em one Higgs rule}, i.e., whenever we turn on a
non-trivial modulus vev and smoothly break some gauge symmetry, then
each gauge boson acquiring a mass is accompanied by one matter field only,
which, of course, plays the r\^ole of the Higgs multiplet. Each
hypermultiplet remains exactly massless unless it delivers the relevant
degrees of freedom for the supersymmetric Higgs effect. As a consequence,
in six dimensions only D-term masses rather than F-term masses are possible.

Let me briefly compare this with the situation in four dimensions. Pure
gravitational anomalies do not exist there, so that a relation like~(\ref{OHR})
cannot in general be deduced. On the other hand, it was observed
in~\cite{IMNQ} that very often in $Z_N$ orbifolds a
{\em three Higgs rule\/} is in effect, which states that each gauge boson
becoming massive is accompanied by three matter fields. This is an
obvious generalization of the above strict one Higgs rule, but there
are cases where it is violated\footnote{I thank Hans Peter Nilles
for an e-mail discussion about this point.}. Nevertheless the existence of
these and other remarkable regularities might result from other kinds of
anomalies.

In order to exploit the anomaly~(\ref{anomaly}) further, we have to use
the following general group theoretical expressions:
\bea
   \Tr       F_A^4 &=&   T_A \tr F_A^4 + U_A   (\tr F_A^2)^2  \\
   \Tr       F_A^2 &=&                   V_A    \tr F_A^2     \\
   \tr_{R^i} F_A^4 &=& t_A^i \tr F_A^4 + u_A^i (\tr F_A^2)^2  \\
   \tr_{R^i} F_A^3 &=&                   w_A^i  \tr F_A^3     \\
   \tr_{R^i} F_A^2 &=&                   v_A^i  \tr F_A^2.
\eea
If the symbol $\tr$ is used without specification it refers to the trace
in the fundamental representation.
Since the coefficients $\mu_A$ in~(\ref{quartic}) have to vanish, we can
write down the following relations:
\be \label{numbgen}
   T_A = \sum_i s_A^i t_A^i \quad \forall A .
\ee
This is the previously mentioned relation which relates the multiplicities
$s_A^i$ to purely group theoretical quantities. If we apply it to
the $E_8 \times E_8$ theory with gauge and spin connection identified
we get an unpleasant surprise. No factor of the resulting
$E_7 \times E_8$ gauge group has a fourth order Casimir invariant.
Thus, relations~(\ref{numbgen}) are trivially satisfied since all
$t_A^i$ vanish. However, if we use the $SO(32)$ theory instead, we
find the required information. In fact, the resulting group is
$SO(28) \times SU(2)$ and for $SO(N)$ groups we have~\cite{GS}
\be \label{SON} \ba{lll}
   \Tr F_{SO(N)}^4 = (N-8)& \tr F_{SO(N)}^4 + 3& (\tr F_{SO(N)}^2)^2, \acht
   \Tr F_{SO(N)}^2 = (N-2)& \tr F_{SO(N)}^2.&
\ea \ee
Hence, $T_{SO(28)} = 20$ so that 20 vector representations are needed to
satisfy condition~(\ref{numbgen}). It is easy to convince oneself that this is
also the number of moduli multiplets\footnote{The simplest way is to take
a four-dimensional point of view upon dimensional reduction and to count the
number of vector representations of $SO(26)$, which are known to correspond to
the number of complex moduli.}, which in turn is given by $h_{1,1}$.

In a similar approach~\cite{Seiberg}, the number of moduli fields could be
determined using the fact that type IIB superstrings can also be compactified
on CY manifolds. In this case, the anomalies from the supergravity multiplet
are cancelled by the dilaton multiplet and 20 further matter multiplets.
In type IIB supergravity theories in ten and six dimensions
(anti)-self-dual antisymmetric tensor fields play an important r\^ole and
contribute to anomalies as the only bosonic fields. The reason for this
is analogous to the reason why chiral fermions contribute. The chiral
respectively (anti)-duality constraints in general give rise to a failure of
gauge and Lorentz invariance at the quantum level~\cite{AW}.
Antisymmetric tensor fields also play a prominent r\^ole in
the Green-Schwarz mechanism. Here, however, the cancellation procedure involves
one-loop {\em and\/} tree diagrams and should not be confused with the
type IIB case, where the anomaly is purely one-loop. In fact, anomaly
freedom for type IIB supergravity in $D=10$ was already observed
in~\cite{AW}; it is not necessary to assume that it arises as the zero slope
limit of a string theory, and no (string motivated) counterterms are needed.

Using the fact that the $SO(28) \times SU(2)$ theory has, like
$E_7 \times E_8$, $y=381$ vector fields we find with help of eq.~(\ref{OHR})
the number of hypermultiplets to be $s=625$. Subtracting the 10 generations
of $({\bf 28},{\bf 2})$, or of ${\bf 56}$ of $E_7$, we are left with
65 singlet states out of which 20 are moduli fields and the remaining 45
are due to $H^1 (End\, T)$.
To summarize, each two-complex-dimensional CY manifold must necessarily
have\footnote{Each hypermultiplet involves two complex scalars.}
\bea
   h_{1,1} = 20, \label{h11} \\
   {\rm dim} H^1 (End\, T) = 90.
\eea
Of course, these equations must also be satisfied in case Gepner's
conjecture\footnote{See footnote on page 1.} fails to be true.

The question arises as to whether it is really necessary to take a detour
in order to arrive at the number of ${\bf 56}$-plets of $E_7$, such as going
to the type IIB string theory or a different vacuum state of the heterotic
string. I will show later in this section that the mere existence of the
hidden $E_8$ in fact gives another proof of eq.~(\ref{h11}).

I proceed by stating another necessary condition for arriving at an anomaly
free result, namely the vanishing of non-Abelian cubic anomalies,
\be
   \sum_{i,j,A,B} s_{AB}^{ij} w_A^i q_B^j = 0 .
\ee
If, for simplicity, we further assume that we have no more than one $U(1)$
factor in the gauge group, we can also drop the last two terms
in expression~(\ref{anomaly}). The remaining part is
\be \ba{rl}
   i(2 \pi)^3 I =&-{1\over 16}[(\tr R^2)^2
 +{1\over 6}(\tr R^2) \sum\limits_A (V_A-\sum\limits_i s^i_A v_A^i)(\tr F_A^2)
 \acht
 &-{2\over 3} \sum\limits_A (U_A-\sum\limits_i s_A^i u_A^i)(\tr F_A^2)^2
 +4 \sum\limits_{i,j,A,B} s_{AB}^{ij} v_A^i v_B^j (\tr F_A^2)(\tr F_B^2)].
\ea \ee
We must require that it can be written in the form
\be \label{anofac}
   i(2 \pi)^3 I=-{1\over 16} [\tr R^2 - \sum\limits_A \alpha_A^{(1)} \tr F_A^2]
     \times     [\tr R^2 - \sum\limits_B \alpha_B^{(2)} \tr F_B^2].
\ee
The coefficients are determined by
\be \label{alp} \ba{ccc}
    \alpha_A^{(1)}+\alpha_A^{(2)} &=&
    {1\over 6} (\sum\limits_{i} s^i_A v_A^i - V_A) =: \tilde{V}_A, \vier
    \alpha_A^{(1)} \alpha_A^{(2)} &=&
    {2\over 3} (\sum\limits_{i} s^i_A u_A^i - U_A) =: \tilde{U}_A,
\ea \ee
yielding
\be
   \alpha_A^{(1,2)} = {\tilde{V}_A\over 2}
                      \pm {1\over 2} \sqrt{\tilde{V}_A^2 - 4 \tilde{U}_A} .
\ee
The non-trivial conditions come along with the cross terms,
i.e.\
\be \label{cross}
  \alpha_A^{(1)} \alpha_B^{(2)} + \alpha_A^{(2)} \alpha_B^{(1)} =
  4 \sum\limits_{i,j} s_{AB}^{ij} v_A^i v_B^j =: \kappa_{A,B} \quad \forall A,B
\ee
must be satisfied identically.
In order to illustrate the above formulae, let me complete the
$SO(28)\times SU(2)$ example. With help of eqs.~(\ref{SON}) we read off
$V_{SO(28)}=26$ and $U_{SO(28)}=3$,
and hence\footnote{See also appendix~\ref{gt}.}
\be
   \tilde{V}_{SO(28)} = -1, \quad\quad\quad \tilde{U}_{SO(28)} = -2,
   \quad\quad\quad \alpha^{(1,2)}_{SO(28)} = 1,-2.
\ee
As for $SU(2)$ we find
\be
   \tilde{V}_{SU(2)} = 46, \quad\quad\quad \tilde{U}_{SU(2)} = 88,
   \quad\quad\quad \alpha^{(1,2)}_{SU(2)} = 2,44.
\ee
Only with this relative assignment of $\alpha^{(1)}$ and $\alpha^{(2)}$ is
condition~(\ref{cross}) satisfied, since we have 10 representations of
$({\bf 28},{\bf 2})$ and $\kappa_{SO(28),SU(2)}=40$. Therefore,
$$ i(2 \pi)^3 I = -{1\over 16}
          [\tr R^2 - {1\over 26}\Tr F^2_{SO(28)} - {1\over 2}\Tr F^2_{SU(2)}]
   \times [\tr R^2 + {1\over 13} \Tr F^2_{SO(28)} - 11\Tr F^2_{SU(2)}], $$
where traces in the adjoint representations of the gauge groups are used.
Note that then the coefficients in the first factor are simply given by
${1\over k_A}$, where $k_A$ is the dual Coxeter number.

Actually this turns out to be a generic feature for groups realized at level~1
Kac-Moody algebras\footnote{On the other hand higher level
string models~\cite{Lewellen} give rise to larger values of $\alpha^{(1)}_A$.
This is in particular true for the model of reference~\cite{RSSS},
where a {\bf 912} representation of $E_7$ is involved.}
and we can write
\be
   \alpha_A^{(1)} = {V_A\over k_A}.
\ee
Thus, with $V_A$ given in appendix~\ref{gt}, it can be shown, that
\be \label{alp1} \ba{lclccl}
   \alpha^{(1)}_{SU(N)}  &=& \alpha^{(1)}_{Sp(N)}  &=& 2, & \vier
   \alpha^{(1)}_{SO(N)}  &=& \alpha^{(1)}_{G_2}    &=& 1, & (N \geq 5) \vier
   \alpha^{(1)}_{F_4}    &=& \alpha^{(1)}_{E_6}    &=& {1\over 3}, & \vier
&& \alpha^{(1)}_{E_7}    &=& {1\over 6}, & \vier
&& \alpha^{(1)}_{E_8}    &=& {1\over 30}.&
\ea \ee
As for $U(1)$ factors one can define a {\em canonical normalization\/} for the
charges (see section~\ref{orbifold}). Then it turns out that
\be \label{alpu1} \ba{lcc}
   \alpha^{(1)}_{U(1)} &=& 1.
\ea \ee

At level~1, I also obtained simple relations\footnote{See appendix~\ref{gt} for
notation.} for $\alpha^{(2)}$:
\be \label{alp2} \ba{lcll}
   \alpha^{(2)}_{SU(N)}  &=& s^{a^{ij}} + (N-4) s^{a^{ijk}} - 2 &(N\geq 4)\vier
   \alpha^{(2)}_{SO(N)}  &=& 2^{(N-6)} s^{\bf 2^{(N-1)}} - 2 & (N \geq 5) \vier
   \alpha^{(2)}_{SU(2)}  &=& {s^{\bf 2} - 16\over 6} & \vier
   \alpha^{(2)}_{SU(3)}  &=& {s^{\bf 3} - 18\over 6} & \vier
   \alpha^{(2)}_{G_2}    &=& {s^{\bf 7} - 10\over 6} & \vier
   \alpha^{(2)}_{F_4}    &=& {s^{\bf 26} - 5\over 6} & \vier
   \alpha^{(2)}_{E_6}    &=& {s^{\bf 27} - 6\over 6} & \vier
   \alpha^{(2)}_{E_7}    &=& {s^{\bf 56} - 4\over 6} & \vier
   \alpha^{(2)}_{E_8}    &=& - {1\over 5}.&
\ea \ee
In the first two cases the number of fundamental representations is fixed
by eq.~(\ref{numbgen}).

The final step in this discussion is the introduction of
a counterterm~\cite{GS} designed to cancel the factorized anomaly.
It can be chosen to be
\be
   \Delta {\cal L}_{GS} = {i\over 16 (2 \pi)^3}
                        B [\tr R^2 - \sum\limits_A \alpha_A^{(2)} \tr F_A^2],
\ee
and cancels the anomaly~(\ref{anofac}) if $B$ transforms according to
\be \label{btrafo}
   B \rightarrow B + [\tr (\omega {\rm d}\Theta) -
                   \sum\limits_B \alpha_B^{(1)} \tr (A_B {\rm d} \Lambda_B )].
\ee
Here $\omega$ and $A_B$ are the Lorentz and gauge connections and
$\Theta$ and $\Lambda_B$ are the respective transformation parameters.

As an example, I will now determine the spectrum of the $E_7\times E_8$ theory:
\begin{enumerate}
\item Since at level~1 there is no matter transforming non-trivially under
      $E_8$, we find $\alpha^{(1)}_{E_8} = 1/30$ confirming~(\ref{alp1}) and
      $\alpha^{(2)}_{E_8} = - 1/5$.
\item Similarly, there are only {\bf 56}-plets of $E_7$ fixing
      $\alpha^{(1)}_{E_7} = 1/6$, again in accordance with~(\ref{alp1}).
\item Since the vector multiplet of $E_8$ is neutral under $E_7$ and there are
      no other non-trivial $E_8$ multiplets, there cannot be mixed $E_7$ and
      $E_8$ gauge anomalies and the associated cross terms in
$$ i(2 \pi)^3 I = -{1\over 16}
          [\tr R^2 - {1\over 6}\tr F^2_{E_7} - {1\over 30}\tr F^2_{E_8}]
   \times [\tr R^2-\alpha^{(2)}_{E_7}\tr F^2_{E_7} + {1\over 5}\tr F^2_{E_8}]$$
      must cancel. Hence, $\alpha^{(2)}_{E_7} = 1$ and
      $s_{E_7}^{56} = 10 = {1\over 2} h_{1,1}$!
\end{enumerate}

{}From a purely field theory point of view we would not allow
for a Green-Schwarz counterterm, since it is not supersymmetric~\cite{GS}.
Consequently, we would require the coefficients $\alpha^{(2)}$ to vanish,
which is clearly possible for all cases (except for $E_8$ at level~1) and would
even give us a non-trivial restriction on the particle content.
However, at level~1,
there is no way of satisfying the finiteness condition\footnote{There are some
solutions for low-dimensional representations at higher level, but they are
certainly not realistic and one has to include a huge number of singlets to
match the constraint~(\ref{OHR}). E.g.\ one could take one symmetric and one
antisymmetric second rank tensor representation of $SU(N)\, (N\geq 3)$.} on
super-Yang-Mills theories~\cite{HSW}. On the other hand, string models in
general have a non-vanishing GS-counterterm and do not respect the
YM-finiteness condition.

\section{$Z_N$ orbifold limits of $K_3$}
\label{orbifold}

In the previous section I have shown that anomaly considerations in six
dimensions lead essentially uniquely to the $K_3$ manifold. I want to use
this fact to show that even asymmetric orbifolds~\cite{NSV}, which seemingly
have no obvious geometrical interpretation since left and right moving
coordinates are twisted in a different way,
are likely to be just singular points on that manifold. Of course,
the orbifold construction presents a distinguished place for studying
anomalies, especially since they typically lead to enhanced gauge groups
and delicate questions concerning $U(1)$ charges and normalizations
can be addressed.

The symmetric orbifold limits of $K_3$ have been discussed in~\cite{Walton},
but as mentioned in the proceeding section, anomaly cancellation conditions
offer an excellent check showing there are some errors for non-prime twists.
Here I include asymmetric orbifolds as well and briefly describe,
how to construct and classify these models.
In an upcoming publication~\cite{asym} I will present the four dimensional
cases completing the list of Abelian $(2,2)$ orbifolds:
Symmetric $Z_N$ orbifolds were discussed in~\cite{DHVW,IMNQ} and updated
and completed in~\cite{EK}. Symmetric $Z_N \times Z_M$ orbifolds including
discrete torsion can be found in~\cite{FIQ}. It is clearly desirable to have
the complete list, since they all correspond to exactly solvable models and
are complements to the Landau-Ginzburg vacua classified in~\cite{KS}.

For each orbifold model we must assure that a number of conditions are met.
Since we are interested in $N=1$ supersymmetric compactifications, we have to
require that only one of the two gravitinos is projected out. Thus the two
right moving internal complex coordinates have to be twisted in the same way.
Likewise, since I want to discuss $(4,4)$ models, this has to be true for
the left movers, as well. On the other hand, we need not insist on treating
left and right movers equally.

The next condition uses the fact that the Lefschetz fixed point theorem
determines the degeneracy in the first twisted sector of an orbifold.
It is given by
\be \label{lef}
   n_1 = |(1-\beta_L)(1-\beta_R)|,
\ee
where the twist eigenvalues $\beta$ are non-trivial $N$th roots of unity.
Since $n_1$ has to be an integer, symmetric $Z_N$ orbifolds
($\beta_L = \beta_R$) can only exist for $N=2,3,4,6$ and $n_1=4,3,2,1$. But
when allowing for $\beta_L \neq \beta_R$, I find two more solutions. They
correspond to $Z_8$ with $n_1=2$ and $Z_{12}$ with $n_1=1$. In fact, there are
precisely two non-trivial 8th and 12th roots of unity (plus there complex
conjugates), one of them corresponding to $\beta_L$ and the other to
$\beta_R$. Thus let me define $\beta_L = e^{2\pi i/8}$, $e^{2\pi i/12}$ and
$\beta_R = e^{6\pi i/8}$, $e^{10\pi i/12}$ for $Z_8$ and $Z_{12}$,
respectively.

It is still necessary to show that these orbifolds actually exist. This can be
done by explicitly constructing the torus lattice in which the twist can act.
As described in~\cite{EJN,thesis} I have to find an order $N$ twist matrix
$\Theta \in O(4,4;{\bf Z})$ acting on winding and momentum quantum numbers
and a background metric $G$ of the
form\footnote{The $4\times 4$ matrices $g$ and $b$ denote constant background
fields; for notation and more details see~\cite{EJN,thesis}.}
\be \label{bigg}
   G = \left( \ba{cc} (g-b)g^{-1}(g+b) & bg^{-1} \\
                      -g^{-1}b & {1\over g} \ea \right),
\ee
such that
\be \label{biggcond}
   [G\Theta - {\Theta^T}^{-1}G] = 0.
\ee

The $Z_{12}$ orbifold can be constructed as the product of two two-dimensional
asymmetric orbifolds with twist matrices
\be
   \Theta_{12} = \left( \ba{cccc}
     0 & 0 & 1 & 0 \\
     0 & 0 & 0 & 1 \\
     1 & 0 & 0 & 1 \\
     0 & 1 &-1 & 0 \ea \right)
\ee
and background fields
\be
g_{12}=\left(\ba{cc}{\sqrt{3}\over 2}&0\\ 0& {\sqrt{3}\over 2}\ea \right),
\quad\quad
b_{12}=\left(\ba{cc}  0   & {1\over 2} \\ -{1\over 2} &    0   \ea \right).
\ee
This two-dimensional $Z_{12}$ model can be shown to be equivalent to the one
discussed in~\cite{HMV}. There it was pointed out that it is an example
of an irrational two-dimensional conformal field theory, which is exactly
solvable and possibly not connected to any rational one. In contrast,
due to supersymmetry in $D=6$ there are many exactly marginal operators
in the twisted sectors; therefore, this model is by no means located
at a single point.

By again referring to the Lefschetz theorem it is clear that a $Z_8$ orbifold
cannot be defined in two dimensions. In four dimensions there are
two possibilities: One is a symmetric orbifold, whose twist matrix can be
written in terms of $4\times 4$ block matrices as
\be \label{z8s}
\Theta_8^S=\left( \ba{cc} \theta_8 & 0 \\ 0 & {\theta_{8}^T}^{-1} \ea \right),
\ee
where
\be
\theta_{8}=\left( \ba{cccc}
       0 & 0 & 0 & -1 \\
       1 & 0 & 0 & 0  \\
       0 & 1 & 0 & 0  \\
       0 & 0 & 1 & 0 \ea \right).
\ee
It cannot, however, lead to space time supersymmetry. The other twist is
asymmetric and given by
\be \label{z8a}
\Theta_8^A=\left( \ba{cc} 0 & \theta_8 \\ {\theta_{8}^T}^{-1} & 0 \ea \right).
\ee
The fact that the twist defined by~(\ref{z8a}) has an asymmetric
form is not enough to ensure that it is not equivalent to the symmetric
one given by~(\ref{z8s}). One way of proving that they do indeed differ, is
to show that the condition~(\ref{biggcond}) can be satisfied only for
a specific background configuration. The metrical background
$g_8$ is just given by the identity matrix and the antisymmetric background
$b_8$ must vanish. In contrast, twist~(\ref{z8s}) allows for deformations
of $g_8$ and non-vanishing $b_8$.

Now let the left movers correspond to the bosonic side of the heterotic
string. The shift vectors acting in the gauge lattice are then given by
$V = {1\over N} (1,1,0^{14})$. The resulting unbroken gauge group is
$E_8 \times E_7 \times U(1)$, except for $Z_2$ where $U(1)$ is enhanced
to $SU(2)$. The $U(1)$ charge can be determined by requiring that all
gauge bosons have to be neutral. Thus we would assign the charge
$Q =c (a_1 + a_2)$ to a state corresponding to a
vector $\vec{S}=(a_1,a_2,\dots,a_{16})$.
The normalization constant $c$ can be determined by noting that in the $Z_2$
case $Q$ would play the r\^ole of the third isospin component.
In that case, following standard conventions, $c=1/2$ and, since
in string theory all gauge couplings are equal at tree level, this is also
the correct normalization for $U(1)$. In other words, the charge of
a state vector $\vec{S}$ is given by $\vec{S} \vec{Q}$ with the charge operator
$\vec{Q} = 1/2(1,1,0^{14})$.

More generally, since {\em all\/} generators
of $E_8 \times E_8$ or $SO(32)$ are normalized in the same way, we find
the correct normalization for any charge operator in any orbifold model to be
\be
   \vec{Q}^2 = {1\over 2}.
\ee
This is the superstring counterpart of the Grand Unification normalization
leading to the prediction of the Weinberg angle.

The spectra can be derived using standard orbifold techniques~\cite{INQ}.
A subtlety occurs for non-prime orbifolds in higher twisted sectors.
Since incorrect results can be found in the literature, I briefly describe
the correct precedure.

Consider for definiteness the second twisted sector of $Z_6$.
One would first determine the spectrum of the first twisted sector of $Z_3$.
After finding the transformation properties of these states w.r.t.\
$Z_6$, one will encounter twist phases $\pm 1$. Non-oscillator states and
in particular all {\bf 56}-states of $E_7$ have phases $+1$. Single oscillators
$\alpha_{-2/3}$, however, contribute a relative minus sign as compared to
double oscillators $\alpha_{-1/3} \alpha_{-1/3}$, both contributing the
same energy. This is because, while the $Z_3$ phase contribution
in both cases is $e^{4\pi i/3}$, the $Z_6$ phase contribution
$e^{\pm 2 \pi i/3}$ is a priori ambiguous. This ambiguity can be resolved
by the observation that $\alpha_{-1/3}$ and $\alpha_{-2/3}$ respectively
correspond to $\alpha_{-1/6}$ and $\alpha_{-5/6}$ in the first sector.
On the other hand, only 5 combinations of the 9 $Z_3$ fixed points are fixed
under $Z_6$, while the other 4 transform with a sign. Clearly, the overall
twist phase must be $+1$, which then determines the degeneracy of the states.

\begin{table}[t]
\label{spectra}
\centering
\begin{tabular}{|c|rl|rl|rl|rl|rl|rl|}
\hline

 && $Z_2$ && $Z_3$ && $Z_4$ && $Z_6$ && $Z_8$ && $Z_{12}$ \\ \hline\hline

   &&({\bf 56},{\bf 2})&&${\bf 56}_{+1/2}$&&${\bf 56}_{+1/2}$&&${\bf
56}_{+1/2}$&& && \\
 U &4&$({\bf 1},{\bf 1})$&2&${\bf 1}_{0}$&2&${\bf 1}_{0}$&2&${\bf 1}_{0}$&& &&
\\
   && &&${\bf 1}_{-1}$&& && && && \\ \hline

   &8&({\bf 56},{\bf 1})&9&${\bf 56}_{-1/6}$&4&${\bf 56}_{-1/4}$&&${\bf
56}_{-1/3}$&
   2&${\bf 56}_{-3/8}$&&${\bf 56}_{-5/12}$ \\
T1 &32&$({\bf 1},{\bf 2})$&45&${\bf 1}_{+1/3}$&24&${\bf 1}_{+1/4}$&8&${\bf
1}_{+1/6}$&
   20&${\bf 1}_{+1/8}$&14&${\bf 1}_{+1/12}$ \\
   && &18&${\bf 1}_{-2/3}$&8&${\bf 1}_{-3/4}$&2&${\bf 1}_{-5/6}$&4&${\bf
1}_{-7/8}$&
   2&${\bf 1}_{-11/12}$ \\
\hline
   && && &5&${\bf 56}_{0}$&5&${\bf 56}_{-1/6}$&3&${\bf 56}_{-1/4}$&&${\bf
56}_{-1/3}$ \\
T2 && && &16&${\bf 1}_{+1/2}$&22&${\bf 1}_{+1/3}$&10&${\bf 1}_{+1/4}$&2&${\bf
1}_{+1/6}$ \\
   && && &16&${\bf 1}_{-1/2}$&10&${\bf 1}_{-2/3}$&6&${\bf 1}_{-3/4}$&2&${\bf
1}_{-5/6}$ \\
\hline
   && && && &3&${\bf 56}_{0}$&2&${\bf 56}_{-1/8}$&2&${\bf 56}_{-1/4}$ \\
T3 && && && &11&${\bf 1}_{+1/2}$&4&${\bf 1}_{+3/8}$&8&${\bf 1}_{+1/4}$ \\
   && && && &11&${\bf 1}_{-1/2}$&4&${\bf 1}_{-5/8}$&4&${\bf 1}_{-3/4}$ \\
\hline
   && && && && &3&${\bf 56}_{0}$&3&${\bf 56}_{-1/6}$ \\
T4 && && && && &9&${\bf 1}_{+1/2}$&12&${\bf 1}_{+1/3}$ \\
   && && && && &9&${\bf 1}_{-1/2}$&6&${\bf 1}_{-2/3}$ \\
\hline
   && && && && && &&${\bf 56}_{-1/12}$ \\
T5 && && && && && &2&${\bf 1}_{+5/12}$ \\
   && && && && && &2&${\bf 1}_{-7/12}$ \\
\hline
   && && && && && &2&${\bf 56}_{0}$ \\
T6 && && && && && &6&${\bf 1}_{+1/2}$ \\
   && && && && && &6&${\bf 1}_{-1/2}$ \\ \hline
\end{tabular}
\caption{$Z_N$ orbifolds in six dimensions. Displayed are the $E_7$ quantum
numbers and $U(1)$ charges ($SU(2)$ quantum numbers in case of $Z_2$).
The states are grouped according to the sector from which they arise.}
\end{table}

All six $Z_N$ orbifold spectra with all appearing $U(1)$ charges are presented
in table~\ref{spectra}. Notice that in the asymmetric cases no matter fields
come from the untwisted sectors. This could have been anticipated, since, as
described above, the background fields $g$ and $b$ are fixed so that they
do not correspond to moduli fields, which would in turn give rise to
{\bf 56}-plets of $E_7$.

In all cases one can verify the charge sum rules
\be \ba{rrr}
   \sum_i Q_i^2 &=& 42, \vier
   \sum_i Q_i^4 &=&  9,
\ea \ee
so that in the notation of section~\ref{ac},
\be
   \tilde{V}_{U(1)} = 7, \quad\quad\quad \tilde{U}_{U(1)} = 6,
   \quad\quad\quad \alpha^{(1,2)}_{U(1)} = 1,6,
\ee
in agreement with~(\ref{alpu1}). Moreover, for {\bf 56}-plets we see that
\be \ba{rrr}
    \sum_i {\bf 56}_{Q_i^2} &=& {\bf 56}_{1/2},
\ea \ee
so that according to equation~(\ref{cross}) the correct mixed $E_7$ and $U(1)$
anomaly occurs, since
\be \label{u1anom}
 i(2 \pi)^3 I = -{1\over 16}
[\tr R^2 - {1\over 6}\tr F^2_{E_7} - {1\over 30}\tr F^2_{E_8} - F^2_{U(1)}]
\times [\tr R^2 - \tr F^2_{E_7} + {1\over 5}\tr F^2_{E_8} - 6 F^2_{U(1)}].
\ee
Finally, the cross terms between $E_8$ and $U(1)$ cancel in~(\ref{u1anom}).

Thus, I could not only derive the number of singlet states with the help of
equations~(\ref{OHR}) and~(\ref{numbgen}), but I also found three
independent conditions which have to be satisfied by the charges.
It should be emphasized that similar stringent sum rules on $U(1)$ charges
can be found in four dimensions as well, regardless of whether the
$U(1)$ is anomalous or not. In fact, the number of such sum rules
increases rapidly with the number of $U(1)$ factors. In fact,
many semi-realistic orbifold models have a large number of $U(1)$'s and
the sum rules can be used, even if most of them are spontaneously broken.

I finally present some orbifold models with non-standard gauge
embeddings. Take the $SO(32)$ model and use the embedding vector
$V = {1\over 3} (1,1,1,1,-2,0^{11})$ for the $Z_3$ orbifold.
The arising matter content transforming under
$SO(22) \times SU(5) \times U(1)_{\tilde{Q}}$ is
\be \ba{rl}
   U: & ({\bf 22},{\bf 5})_{+1/2} + ({\bf 1},\overline{\bf 10})_{-1} +
     2\,({\bf 1},{\bf 1})_0, \vier
   T: & 9\,({\bf 22},{\bf 1})_{+5/6} + 18\,({\bf 1},\overline{\bf 5})_{+1/3} +
        9\,({\bf 1},{\bf 10})_{-2/3}.
\ea \ee
The properly normalized charge is given by $Q=\sqrt{2\over 5} \tilde{Q}$.
I mention this model to show how the cubic and the non-factorizable quartic
anomalies are cancelled between the untwisted and twisted sector.

The other examples use the twist embedding formulation~\cite{DHVW}, which
permits {\em continuous\/} Wilson lines breaking the rank of the gauge
group~\cite{INQ2}. This is an explicit realization of the flat directions
discussed in~\cite{FINQ}. Flat directions along Wilson lines
are especially interesting, because they do not lead
out of the orbifold class and can be studied in great detail. For instance,
it is a unique place to derive the modular group of the corresponding
$(0,2)$ moduli, using methods developed in~\cite{EJN,thesis,ES}.

Consider the $Z_3$ orbifold with standard twist embedding. The twist acts as
a rotation in one factor of the $SU(3)^4$ subgroup of $E_8$ and is equivalent
to the standard shift embedding. The spectrum is given in
table~\ref{spectra}. In this formulation there are four smooth Wilson line
directions. Turned on, they break $E_7 \times U(1)$ to $E_6$. In this process
the untwisted {\bf 56} and the untwisted charged singlet become massive as the
longitudinal parts of the ${E_7 \times U(1)\over E_6}$ gauge bosons except
for one $E_6$ neutral combination, which becomes a massless $(0,4)$ modulus
multiplet, whose scalar components correspond to the Wilson line directions.
Thus we arrive at a model with 18 {\bf 27}-plets of $E_6$.

Next consider the non-standard embedding case, where the $Z_3$ rotation acts
in all the four $SU(3)$ subgroups simultaneously. This yields an
$SU(9)\times E_8$ gauge group with matter content
\be \ba{rl}
   U: & {\bf 84} + 2\,{\bf 1}, \vier
   T: & 9\,{\bf 36} + 18\,{\bf 9}.
\ea \ee
In this case, the {\bf 84} serves as the Higgs representation breaking $SU(9)$
completely and also leaving four moduli fields. Thus, starting with a quite
non-trivial model, the flat directions ``trivialize'' it leaving only a
pure $E_8$ YM theory with 486 singlets and 6 (untwisted) moduli.

Finally, it is possible to twist three $SU(3)$ subgroups of the second $E_8$
as well. I find the gauge group $SU(9)\times E_6\times SU(3)$ with matter
transforming as
\be \ba{rl}
   U: & ({\bf 84},{\bf 1},{\bf 1}) + ({\bf 1},{\bf 27},{\bf 3}) +
     2\,({\bf 1},{\bf 1},{\bf 1}), \vier
   T: & 9\,({\bf 9},{\bf 1},{\bf 3}).
\ea \ee
The generic gauge group after Wilson line breaking, however, is just
$SU(3)$, and we are left with 81 triplets and 9 (untwisted) moduli.
Further symmetry breaking generally occurs after turning on twisted
flat directions and we are faced with the situation that the stringy Higgs
effect seems to be too efficient. Generally we are left without any
charged matter, and the same is true in four dimensions.

String model building so far has been concentrated on finding the gauge group
of the standard model or some unifying group. Additional
$U(1)$'s were broken by using mechanisms such as the one just described.
However, given the observations above, the fact that some charged fields
(the observed quarks and leptons) remain in the massless spectrum appears as a
fine tuning if there are further flat directions breaking, e.g.
$U(1)_{\rm EM}$. One possibility is to find a reason as to why the
unwanted flat directions are not ``used'' by nature. Of course, that
would require a much better understanding of string dynamics. It seems
much more likely, however, that the solution to the above problem
can be found in the non-existence of further flat direction. If this
is true, it would be very important to find examples where charged matter
remains at a generic point in moduli space. It might even turn out that some
of the models presented in the literature possess this property. However,
to my knowledge, they have not yet been examined in this respect.

It should be possible to trace back the non-existence of certain flat
directions to some kind of index and/or anomaly. Gauge and gravitational
anomalies are certainly not sufficient here. If such an index could be
found, this would offer an excellent phenomenological opportunity for
discarding a large number of string models which lack the protecting index,
like the six-dimensional examples above.

\section*{Acknowledgements}
It is a pleasure to thank Mirjam Cveti$\check{\rm c}$ and Albrecht Klemm for
many valuable discussions as well as Ramy Brustein and Paul Langacker for
helpful suggestions and comments.

\appendix
\section{Group theoretical identities}
\label{gt}

Relations between adjoint and fundamental representations of the classical
Lie groups were derived in~\cite{PVN}. This appendix is meant to be a more
extended reference and includes spinor and third rank antisymmetric tensor
representations as well as all exceptional groups.

Relations defining $V_A$:
\be \label{va} \ba{lccl}
   \Tr F_{SU(N)}^2 &=& 2N    &\tr F_{SU(N)}^2 \vier
   \Tr F_{SO(N)}^2 &=& (N-2) &\tr F_{SO(N)}^2 \vier
   \Tr F_{Sp(N)}^2 &=& (N+2) &\tr F_{Sp(N)}^2 \vier
   \Tr F_{G_2}^2   &=& 4     &\tr F_{G_2}^2   \vier
   \Tr F_{F_4}^2   &=& 3     &\tr F_{F_4}^2   \vier
   \Tr F_{E_6}^2   &=& 4     &\tr F_{E_6}^2   \vier
   \Tr F_{E_7}^2   &=& 3     &\tr F_{E_7}^2   \vier
   \Tr F_{E_8}^2   &=&       &\tr F_{E_8}^2.
\ea \ee
The last equation serves as a definition\footnote{In~\cite{GSW}
the definition $\Tr F_{E_8}^2 = 30\, \tr F_{E_8}^2$ is used.}.

Relations defining $T_A$ and $U_A$:
\be \label{taua} \ba{lcclccl}
   \Tr F_{SU(N)}^4 &=&   2N  &\tr F_{SU(N)}^4& +& 6 &(\tr F_{SU(N)}^2)^2 \vier
   \Tr F_{SO(N)}^4 &=& (N-8) &\tr F_{SO(N)}^4& +& 3 &(\tr F_{SO(N)}^2)^2 \vier
   \Tr F_{Sp(N)}^4 &=& (N+8) &\tr F_{Sp(N)}^4 &+& 3 &(\tr F_{Sp(N)}^2)^2 \vier
   \Tr F_{G_2}^4   &=&       &     & &   {5\over 2} &(\tr F_{G_2}^2)^2   \vier
   \Tr F_{F_4}^4   &=&       &     & &  {5\over 12} &(\tr F_{F_4}^2)^2   \vier
   \Tr F_{E_6}^4   &=&       &     & &   {1\over 2} &(\tr F_{E_6}^2)^2   \vier
   \Tr F_{E_7}^4   &=&       &     & &   {1\over 6} &(\tr F_{E_7}^2)^2   \vier
   \Tr F_{E_8}^4   &=&       &     & & {1\over 100} &(\tr F_{E_8}^2)^2.
\ea \ee

For groups with an independent fourth order Casimir invariant, by definition
$v_A^f=t_A^f=1$ and $u_A^f=0$ for fundamental representations $f$.
Otherwise, $t_A^f=0$ and $u_A^f$ can be extracted from
\be \label{fund} \ba{lccl}
   \tr F_{SU(2)}^4 &=& {1\over 2}   &(\tr F_{SU(2)}^2)^2 \vier
   \tr F_{SU(3)}^4 &=& {1\over 2}   &(\tr F_{SU(3)}^2)^2 \vier
   \tr F_{G_2}^4   &=& {1\over 4}   &(\tr F_{G_2}^2)^2   \vier
   \tr F_{F_4}^4   &=& {1\over 12}  &(\tr F_{F_4}^2)^2   \vier
   \tr F_{E_6}^4   &=& {1\over 12}  &(\tr F_{E_6}^2)^2   \vier
   \tr F_{E_7}^4   &=& {1\over 24}  &(\tr F_{E_7}^2)^2   \vier
   \tr F_{E_8}^4   &=& {1\over 100} &(\tr F_{E_8}^2)^2.   \vier
\ea \ee

Other representations, which appear in string models realized
at level~1 w.r.t.\ the underlying Kac-Moody-algebra, are totally antisymmetric
tensor representations of higher rank for $SU(N)$ and the lowest dimensional
spinor representations of $SO(N)$. The second resp.\ third relations
in~(\ref{va}) and~(\ref{taua}) serve as formulae for antisymmetric resp.\
symmetric second rank tensor representations for all the classical Lie groups.

In order to fix $w_{SU(N)}^{a^{ij}}$ for second rank antisymmetric tensor
representations $a^{ij}$, I note
\be \ba{lccl}
   \tr_{a^{ij}} F_{SU(N)}^3&=&(N-4)&\tr F_{SU(N)}^3 \quad\quad\quad (N \geq 3).
\ea \ee

As for third rank totally antisymmetric tensor representations $a^{ijk}$
it can be shown that
\be \label{trtr} \ba{lcclccl}
   \tr_{a^{ijk}} F^2 &=& {1\over 2}(N^2 - 5 N + 6)&\tr F^2 &&& \vier
   \tr_{a^{ijk}}F_{SU(N)}^3&=&{1\over 2}(N^2-9N+18)&\tr F_{SU(N)}^3 &&& \vier
   \tr_{a^{ijk}} F^4 &=& {1\over 2}(N^2-17N +54)&\tr F^4&+&(3N-12)&(\tr F^2)^2.
\ea \ee
Note that in particular the right hand sides vanish for $N=3$ and that
$$\tr_{a^{ijk}} F_{SU(4)}^4 = \tr F_{SU(4)}^4, \quad\quad
\tr_{a^{ijk}} F_{SU(4)}^2 = \tr F_{SU(4)}^2 \quad {\rm and} \quad
\tr_{a^{ijk}} F_{SU(4)}^3 = -\tr F_{SU(4)}^3,$$
as is expected, since $a_{SU(3)}^{ijk}={\bf 1}$ and
$a_{SU(4)}^{ijk}={\bf {\bar 4}}$.

For the basic spinor representations I find,
\be \ba{lcrlcccl}
   \tr_{2^{(N-1)}} F_{SO(2N)}^2 &=& 2^{(N-4)} &\tr F_{SO(2N)}^2 &&&& \vier
   \tr_{2^{(N-1)}} F_{SO(2N)}^4 &=& -2^{(N-5)} &\tr F_{SO(2N)}^4
                                &+&3&2^{(N-7)}&(\tr F_{SO(2N)}^2)^2,
\ea \ee
and these relations continue to hold after replacement
of $SO(2N)$ with $SO(2N-1)$.

I end this appendix by showing how relations for exceptional groups can be
found. For instance, the last relation in~(\ref{fund}) can
be proved as follows:
Under $SU(9)$ the fundamental ${\bf 248}$ of $E_8$ has the decomposition
\be
   {\bf 248} \rightarrow {\bf 80} + {\bf 84} + {\bf \overline {84}}.
\ee
For the reducible {\bf 248} representation of $SU(9)$ I find, with help
of eqs.~(\ref{taua}) and (\ref{trtr}),
$$\tr_{248} F^4_{SU(9)} = \tr_{80} F^4_{SU(9)} + \tr_{84} F^4_{SU(9)} +
                             \tr_{\overline {84}} F^4_{SU(9)}$$
$$   = 18\,\tr F^4_{SU(9)} + 6\, (\tr F^2_{SU(9)})^2
     +2\, [-9\,(\tr F^2_{SU(9)})^2 +15\, \tr F^4_{SU(9)}]
     = 36\, (\tr F^2_{SU(9)})^2.$$
Note, how the independent forth order invariant cancels out. Similarly,
$$\tr_{248} F^2_{SU(9)} = 60\, \tr F^2_{SU(9)},$$
and so
\be
   {\tr_{248} F^4_{SU(9)}\over (\tr_{248} F^2_{SU(9)})^2} =
   {\tr F^4_{E_8} \over (\tr F^2_{E_8})^2} = {1\over 100}.
\ee


\begin{thebibliography}{99}
\bibitem{GS} M. B. Green and J. H. Schwarz, {\em Phys. Lett.}
             {\bf 149B} (1984) 117.
\bibitem{AW} L. Alvarez-Gaum\'e and E. Witten, {\em Nucl. Phys.} {\bf B234}
             (1983) 269.
\bibitem{tHooft} G. 't Hooft, in {\em Recent Developments in Gauge Theories},
                 G. 't Hooft et. al. eds., Plenum, New York (1980).
\bibitem{CHSW} P. Candelas, G. T. Horowitz, A. Strominger and E. Witten,
               {\em Nucl. Phys.} {\bf B258} (1985) 46.
\bibitem{Seiberg} N. Seiberg, {\em Nucl. Phys.} {\bf B303} (1988) 286.
\bibitem{KS} M. Kreuzer and H. Skarke, {\em Nucl. Phys.} {\bf B388} (1992)
             113; \\
             A. Klemm and R. Schimmrigk, {\em Landau-Ginzburg String Vacua},
             preprint HD-THEP-92-13 (TUM--TH--142/92, CERN--TH--6459/92).
\bibitem{NSV} K. S. Narain, M. H. Sarmadi and C. Vafa, {\em Nucl. Phys.}
              {\bf B288} (1987) 551.
\bibitem{Gepner} D. Gepner, {\em Phys. Lett.} {\bf 199B} (1987) 380.
\bibitem{DHVW} L. Dixon, J. A. Harvey, C. Vafa and E. Witten, {\em Nucl. Phys}
               {\bf B261} (1985) 678; {\em Nucl. Phys.} {\bf B274} (1986) 285.
\bibitem{IMNQ} L. E. Ib\'a\~nez, J. Mas, H. P. Nilles and F. Quevedo,
               {\em Nucl. Phys.} {\bf B301} (1988) 157.
\bibitem{EK} J. Erler and A. Klemm, {\em Comment on the Generation Number in
             Orbifold Compactifications}, Munich preprint MPI--Ph/92-60
             (TUM--TH--146/92), to appear in {\em Comm. Math. Phys.}
\bibitem{asym} J. Erler, in preparation.
\bibitem{Walton} M. A. Walton, {\em Phys. Rev.} {\bf D37} (1988) 377.
\bibitem{FT} S. Ferrara, D. L\"ust, A. Shapere and S. Theisen,
             {\em Phys. Lett.} {\bf 225B} (1989) 363; \\
             S. Ferrara and S. Theisen,
             {\em Moduli Spaces, Effective Actions and Duality Symmetry in
             String Compactifications}, CERN preprint TH.5652/90,
             (UCLA/90/TEP/8), lectures given at the 3rd Hellenic Summer School,
             Corfu, Greece, Sep. 13--23, 1989.
\bibitem{GSWest} M. B. Green, J. H. Schwarz and P. C. West,
                 {\em Nucl. Phys.} {\bf B254} (1985) 327.
\bibitem{GSW} M. B. Green, J. H. Schwarz and E. Witten, {\em Superstring
              Theory}, Vol. 2, Cambridge University Press, Cambridge (1987).
\bibitem{RSSS} S. Randjbar-Daemi, A. Salam, E. Sezgin and J. Strathdee,
               {\em Phys. Lett.} {\bf 151B} (1985) 351.
\bibitem{Lewellen} D. C. Lewellen, {\em Nucl. Phys.} {\bf B337} (1990) 61.
\bibitem{HSW} P. Howe, K. Stelle and P. C. West, {\em Phys. Lett.} {\bf 124B}
              (1983) 55.
\bibitem{FIQ} A. Font, L. E. Ib\'a\~nez and F. Quevedo,
              {\em Phys. Lett.} {\bf 217B} (1989) 272.
\bibitem{EJN} J. Erler, D. Jungnickel and H. P. Nilles, {\sl Phys. Lett.}
              {\bf 276B} (1992) 303.
\bibitem{thesis} J. Erler, {\em  Investigation of Moduli Spaces in String
                 Theories}, thesis at the Technical University of Munich
                 (in German language), preprint MPI-Ph/92-21.
\bibitem{HMV} J. Harvey, G. Moore and C. Vafa, {\em Nucl. Phys.} {\bf B304}
              (1987) 269.
\bibitem{INQ} L. E. Ib\'a\~nez, H. P. Nilles and F. Quevedo, {\em Phys. Lett.}
              {\bf 187B} (1987) 25.
\bibitem{INQ2} L. E. Ib\'a\~nez, H. P. Nilles and F. Quevedo, {\em Phys. Lett.}
               {\bf 192B} (1987) 332.
\bibitem{FINQ} M. Cveti$\check{\rm c}$, {\em Phys. Rev. Lett.} {\bf 59} (1987)
               2829; \\
               A. Font, L. E. Ib\'a\~nez, H. P. Nilles and F. Quevedo,
               {\em Nucl. Phys.} {\bf B307} (1988) 109.
\bibitem{ES} M. Spali\'{n}ski, {\em Nucl. Phys.} {\bf B377} (1992) 339; \\
             M. Spali\'nski, {\em Phys. Lett.} {\bf 275B} (1992) 47; \\
             J. Erler and M. Spali\'nski, {\em  Modular Groups for Twisted
             Narain Models}, Munich preprint MPI--Ph/92--61 (TUM-TH-147/92).
\bibitem{PVN} P. van Nieuwenhuizen, {\em Anomalies in Quantum Field Theory:
              Cancellation of Anomalies in d=10 Supergravity},
              Leuven University Press, Leuven (1988).
\end{thebibliography}
\end{document}